 \RequirePackage{amsthm}
\documentclass[pdflatex,referee, sn-mathphys-num]{sn-jnl}

\usepackage{graphicx}%
\usepackage{multirow}%
\usepackage{amsmath,amssymb,amsfonts}%
\usepackage{amsthm}%
\usepackage{mathrsfs}%
\usepackage[title]{appendix}%
\usepackage{xcolor}%
\usepackage{textcomp}%
\usepackage{manyfoot}%
\usepackage{booktabs}%
\usepackage{algorithm}%
\usepackage{algorithmicx}%
\usepackage{algpseudocode}%
\usepackage{listings}%
\usepackage{pdfpages}
\usepackage{ulem}
\usepackage{xcolor}
\usepackage{soul}
\usepackage{caption}
\usepackage{subcaption, graphicx}
\usepackage{anyfontsize}

\theoremstyle{thmstyleone}%
\theoremstyle{thmstyletwo}%

\theoremstyle{thmstylethree}%

\begin{document}

\title[Article Title]{Integrating Explainable AI in Medical Devices: Technical, Clinical and Regulatory Insights and Recommendations}


\author[1]{\fnm{Dima} \sur{Alattal}}\email{dima.alattal@brunel.ac.uk}

\author[2]{\fnm{Asal} \sur{Khoshravan Azar}}\email{asal.khoshravanazar@brunel.ac.uk}

\author[3,4]{\fnm{Puja} \sur{Myles},
\fnm{Richard} \sur{Branson}, 
}\email{puja.myles@mhra.gov.uk, richard.branson@mhra.gov.uk} \equalcont{These authors contributed equally to this work.}

\author[5]{\fnm{Hatim} \sur{Abdulhussein}}\email{hatim.abdulhussein@nhs.net}

\author*[6]{\fnm{Allan} \sur{Tucker}}\email{allan.tucker@brunel.ac.uk}

\affil*[1,2,6]{\orgdiv{Computer Science Department}, \orgname{Brunel University London}, \orgaddress{ \country{UK}}}

\affil[3,4]{ \orgname{Medicine and Healthcare products Regulatory Agency}, \country{UK}}

\affil[5]{ \orgname{NHS England}, \country{UK}}




 \abstract{\textbf{Purpose: }There is a growing demand for the use of Artificial Intelligence (AI) and Machine Learning (ML) in healthcare, particularly as clinical Decision Support Systems (CDSS) to assist medical professionals. However, the complexity of many of these models, often referred to as black box models, raises concerns about their safe integration into clinical settings as it is difficult to understand how they arrived at their predictions. Explainable Artificial Intelligence (XAI) offers a potential solution by providing justifications for the decisions produced by these models, thereby enhancing trust and understanding among clinicians. To address the aspects of trust and safety, it is essential to consider AI medical devices from various perspectives, including clinical, technical, and regulatory perspectives.
\\
 \textbf{Methods:} This paper discusses insights and recommendations derived from an expert working group convened by the UK Medicines and Healthcare products Regulatory Agency (MHRA). The group consisted of healthcare professionals, regulators, and data scientists, with a primary focus on evaluating outputs from different AI/XAI algorithms in clinical decision-making contexts. Additionally, the group evaluated findings from a pilot study investigating clinicians' behavior and interaction with XAI methods during clinical diagnoses.
\\
 \textbf{Results:} While the data science team provides technical results, the regulators and clinicians highlight their main concerns and recommendations for using AI/XAI methods in CDSS. The study reveals an overall increase in clinicians' trust and diagnostic accuracy when using local explanations, although over-reliance on AI suggestions raises safety concerns from a legal perspective. The study also underscores the importance of other explanation methods in clinical settings from different perspectives, such as global and counterfactual explanations.
\\
 \textbf{Conclusion:} Incorporating XAI methods is crucial for ensuring the safety and trustworthiness of medical AI devices in clinical settings. Adequate training for stakeholders is essential to address potential issues, and further insights and recommendations for safely adopting AI systems as CDSS are provided. }
 



\maketitle
\section{Introduction} \label{sec1}
\subsection{Background}
In clinical settings, AI/ML models can be used as decision support systems through supporting healthcare providers and automating routine tasks\cite{1}. These systems assist clinicians in diagnosing diseases and making decisions about treatment. Unlike conventional CDSS, which match patient characteristics to an existing knowledge base, AI/ML based CDSS (AI-CDSS) apply models trained on data from patients with similar conditions.
Despite its potential, AI is not a universal solution and brings novel questions and significant challenges. Some are related to the nature of AI models and others are related to regulatory, medical, and patient perspectives\cite{52}. Therefore, a multidisciplinary assessment is essential for the safe introduction of AI-based medical devices in clinical settings. 

Clinicians trust in AI-CDSSs is crucial for a safe implementation of AI in clinical settings. However, it can be challenging to build trust in these systems in a setting where clinicians are required to make urgent decisions that could have serious consequences\cite{1}. XAI has emerged as a potential method for the safe implementation of AI. Accurate diagnosis alone may not suffice; an explanation for the AI's decision-making process is crucial \cite{53}. This necessity for transparency was highlighted from the very early days of AI diagnostic systems, where researchers found that the ability to explain decisions was the most desired feature among clinicians\cite{54}. Recent studies corroborate this, showing that clinicians value understanding the reasoning behind complex AI systems, often referred to as "black boxes", particularly when their recommendation do not align with clinical expectations \cite{52,65}. Black box models that often use millions of parameters to capture the non-linearity of input features is a major challenge as it undermines trust and hinders interpretation of the models' predictions. Research indicates that XAI can enhance transparency and trust, potentially leading to better healthcare outcomes even if the diagnostic accuracy is not perfect \cite{53}.

XAI in general refers to the characteristic of an AI-driven system that allows a person to understand the reasoning for a model's outputs. XAI aims to provide interpretability, explainability, and transparency to support healthcare practitioners in their decision-making. Interpretability involves comprehending the inner workings of the model and understanding how it generates predictions. On the other hand, explainability focuses on providing clear and understandable explanations for particular AI decisions, actions, or recommendations. Transparency in the context of XAI, ensures that all stakeholders have a clear understanding of the functioning of an AI system. This could involve for example providing stakeholders with information about the data used, how it is processed, and the underlying assumptions guiding the development of the AI system. Nevertheless, the debate around XAI extends beyond technical aspects, touching on regulatory and ethical concerns that could impede progress if not adequately addressed. Without thorough consideration of XAI methods, AI technologies might neglect core ethical and professional principles, overlook regulatory concerns, and cause significant harm \cite{55}. Therefore, XAI is expected to enhance decision confidence, generate hypotheses about causality, and increase trustworthiness and acceptability of the system. It could also help uncover historical actions and biases embedded in AI models trained with historical data \cite{5}. More investigation is needed to ensure healthcare professionals can understand and rely on XAI methods in CDSSs\cite{56}. This work adopts a multidisciplinary view to explore how XAI can facilitate the safe introduction of AI in clinical settings, emphasising the importance of this feature for building clinicians trust and ensuring regulatory compliance.
This paper reports the findings of the group, focusing on:
\begin{itemize}
    \item Identifying key concerns and recommendations of regulators regarding using different AI/ML models that vary in their complexity in AI-CDSS.
    \item Investigating the information needs and main concerns of medical professionals when employing XAI CDSSs in daily clinical situations.
    \item Evaluating state-of-the-art explanation methods for providing meaningful and helpful explanations in clinical settings.
    \item Providing a set of recommendations and insights to guide the adoption of AI/XAI in clinical settings.   
\end{itemize}

 \subsection{Clinical and Regulatory Perspectives}

AI systems used in healthcare are deemed medical devices by the MHRA in the UK if they are designed for medical purposes like diagnosis, treatment, or monitoring. As a result, deploying these devices requires navigating a complex regulatory and ethical framework to ensure they comply with safety, quality, and performance standards. As AI continues to evolve at a rapid rate, the risks associated with not fully utilising its capabilities are becoming more concerning \cite{new2}. A primary challenge for AI/ML as a medical device is there is ambiguity in applying clinical evidence requirements and evaluating the performance and effectiveness of these models\cite{57}.

A major area to consider when evaluating the effectiveness of the systems is understanding that the link between a specific belief in automation and its actual capabilities varies. These Trust-related biases are the main factor behind algorithm trust (algorithm appreciation) ,overtrust and distrust (algorithm aversion) \cite{66, 68}. Trust calibration in AI refers to the process of appropriately adjusting the level of trust that human users have in AI systems based on their actual reliability\cite{8}. It is important for human-AI collaboration to have trust that is properly calibrated to ensure safety and efficiency\cite{9}. Poorly calibrated trust in CDSS can lead to serious issues with safety \cite{10}. Trust calibration involves understanding the limitations and likely failures of AI systems and adjusting trust in their outputs accordingly\cite{11}.

Explanation of AI predictions is thought to be a solution for this problem and is considered a prerequisite for medical AI \cite{12}. However, some argue that trust calibration errors can also occur when users interact with explanations provided by AI systems, leading to irrational or ill considered agreement or disagreement with AI recommendations\cite{13}. As a result, practitioners should retain the authority to make final decisions to ensure that AI systems are effective at diagnosing and treating patients\cite{59}.

XAI allows practitioners to make more nuanced and informed decisions. However, to ensure effectiveness it is important to evaluate the human–AI interaction. Observing how end-users understand explanations and evaluating the explanations influence on user behaviour requires engaging end-users such as healthcare providers, and assessing their usage in decision-making contexts\cite{2}.

“Assurance” in AI-CDSS refers to the confidence that a system will behave as intended in its intended environment, with a focus on patient safety. Assurance involves both verification and validation: verification ensures that the system is built correctly according to the defined requirements, while validation ensures that the right system is built, meeting the intended purpose and goals. In situations where requirements are implicit, XAI methods are used to provide explanations that allow for direct validation of the ML model. These explanations show that the predictions are based on reliable clinical variables and are consistent with clinical knowledge\cite{14}. 

When evaluating the CDSS performance, robustness is considered an important factor, which refers to the model's ability to maintain its performance even when input features vary slightly. Since safety in an AI/ML application is dependent on explainability and performance, and there is no binary choice between them, safety requirements should be partially, even if not entirely, transformed into explainability requirements\cite{2}. Therefore, if an interpretable model can achieve performance levels comparable to a black box model, the interpretable model should be preferred \cite{14,23}. 

\subsection{Technical Perspective for XAI}
There is a debate about the trade-off between a model's performance and its interpretability. AI/ML models with higher performance tend to be based on more complex algorithms, which can make them less interpretable which in turn makes it difficult to understand how they arrive at their predictions\cite{15,16} .Conversely, models with greater interpretability, commonly known as white-box models, may compromise some performance to deliver transparent and understandable outputs\cite{17}. 

The challenge lies in finding a balance between the two, where the model is more accurate and understandable. XAI methods have been proposed to address this issue by producing human-interpretable representations of ML/AI models. These methods can contribute to safety assurance in healthcare by providing evidence to support the safety of complex AI/ML-based systems\cite{19}. However, this trade-off is not always gradual and can vary depending on the specific application\cite{62,18}. 

XAI can be achieved using intrinsically interpretable models which are models that are transparent and explainable by design or through post-hoc XAI methods that provide explanations without opening the complex black box model \cite{21,22}. Model-agnostic XAI methods refer to techniques that provide explanations for the output of AI systems without relying on the internal workings of the specific AI/ML models used \cite{23}. These explanations can be provided at both local and global levels, highlighting the contribution of different features to the model's output \cite{25}. Local explanations explain an individual decision based on one case, supporting trust in that individual decision, while global explanations, explain a model more generally across the entire training set, thus ensuring that the model behaves reasonably when deployed\cite{26}. In clinical settings, both local and global explanations are highly relevant, as they align with the methods clinicians use to justify their diagnoses and treatments. Clinicians often explain how a disease or diagnostic process works in general (global explanation) and justify specific diagnoses based on individual patient data, such as symptoms, test results, and medical history (local explanation)\cite{58}. This parallel enhances the potential for XAI to support clinical decision-making effectively.

XAI methods offers explanations in various forms such as natural text, parameter influence and data visualisations\cite{6}. The use of visualizations by XAI systems enhances the transparency and comprehensibility of decisions, although clinicians' preferences for explanation methods and types vary significantly and often differ from those of developers \cite{67}. The Local Interpretable Model-agnostic Explanations (LIME) is a well-known method for providing local explanations based on which individual features impact a decision \cite{26,27}. LIME uses a model-agnostic and application-agnostic approach to extract explanations from AI models regardless of domain. However, this comes with a drawback that model-agnostic approaches cannot always meet the specific user requirements and the domain-appropriate explanations\cite{28, 69}.  \\
Counterfactual explanations are a particular type of explanation which relates what may have occurred if a model's input had been altered in a particular way. Users receive practical feedback that they can employ to modify their features and move towards the desirable side of the decision boundary. Unlike other XAI techniques, they offer recommendations on how to get the desired result rather than directly addressing \textit{why} the model made a particular prediction\cite{51}. This approach is particularly valuable in clinical settings, as it can help clinicians and patients understand how to alter risk factors to potentially reverse adverse health probabilities. Additionally, counterfactual explanations can enhance trust, as they allow users to familiarise themselves with unknown processes by understanding the hypothetical input conditions under which the output changes \cite{new1}.

\section{Stage 1: AI/XAI methods evaluation in the workshops} 
The study involved a series of four structured workshops, organised around three distinct themes: regulatory, clinical, and data analytic considerations. The regulatory and clinical themed discussions focused on the necessary level of transparency from both regulatory and clinical perspectives, while considering which information and metrics were particularly valuable for regulators and clinicians. The data analytic considerations were informed by data science experiments generating decisions with explanations using various modeling approaches in response to regulatory requirements. 
In the initial phase of the experiment, the objective was to present performance metrics for a variety of representative white-box and black-box models, such as logistic regression, random forest, and artificial neural networks (ANN), to the expert group. Interpretations were provided for the white-box models, while global and local explanations were provided for a black-box model. Specifically, odds ratios, mean decrease GINI, and permutation importance as were presented as global explainers, LIME as a local explainer, and Explainable Matrix (ExMatrix) for identifying counterfactual cases. The findings of these techniques were provided to the expert group to to obtain their views on using these strategies in CDSSs.

\subsection{Data} 
The chosen task was to predict whether individuals were at a high or low risk of having a heart attack. A high-fidelity synthetic dataset based on anonymized CPRD primary care data was used for this purpose \cite{29}.This synthetic dataset focused on cardiovascular disease risk factors and included 22 variables, such as smoking behavior, age, and chronic conditions associated with cardiovascular disease, for 10,000 synthetic individuals randomly sampled from the dataset. The target variable was binary, confirming or denying the occurrence of a heart attack.

\subsection{Methods} \label{sec:methods}
\subsubsection{AI/ML Models}

\begin{figure}[htbp]
    \centering
    \includegraphics[scale=0.9]{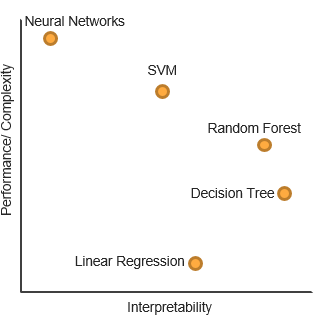}
    \caption{Tradeoff between Performance and Complexity of the model Vs Perceived Interpretabilty in Machine Learning}
    \label{fig:modelscomplex}
    \end{figure}

The expert group selected three ML/AI models to predict patients' risk of having a heart attack: logistic regression, random forest, and neural networks. This study is primarily concerned with assessing the performance of these models and investigating several XAI methods that can be applied to them. The selection of models aimed to cover varying degrees of interpretability and complexity—low, moderate, and high- as outlined in \citep{62} and shown in Figure \ref{fig:modelscomplex}. We explored a number of parameterisations and train-test regimes to assess model performance. The results documented come from a 10-fold cross validation to reduce the risk of overfitting and improve robustness. The expert working group was provided with a mini-tutorial in the form of a presentation, designed to help them interpret the machine learning results. This tutorial introduced the machine learning models discussed in the workshop and aimed to demonstrate their complexity to the panel. It was presented to the group collectively, rather than individually. Although the experts came from varied backgrounds, they had sufficient familiarity with machine learning in healthcare to grasp the material. The tutorial also covered both local and global explanation methods, explaining how they are ranked and what key metrics, such as mean decrease Gini and odds ratios, represent.

\textbf{Logistic regression}\\
Logistic regression is used for binary classification by modeling the probability of a given outcome based on one or more predictor variables. It estimates the relationship between the predictors and the log-odds of the outcome, enabling the prediction of categorical outcomes ans it considered the simplest model in the study \cite{63}.

\textbf{Random forest} \\
A random forest is an ensemble machine learning algorithm that enhances the robustness of decision trees. At its core, the random forest operates by constructing a multitude of decision trees during the training phase. These trees are grown using random subsets of both data and features, injecting an element of randomness into each tree's construction. During prediction, the random forest aggregates the outputs of all constituent trees, typically by taking a majority vote in classification tasks or averaging predictions in regression tasks. As a result, the ensemble approach improves the model's overall accuracy and robustness, while maintaining the interpretable nature of decision trees \cite{63}.

\textbf{Neural Networks} \\
ANN in its most general form consists of layered structures. These layers of interconnected nodes form a network that processes data in stages. The input layer receives the original data and then passes it through hidden layers before arriving at the final prediction in the output layer. 
Researchers have used ANN models as classifiers for risk prediction in the domain of medical diagnostics such as in \cite{36,37}. 

Simple ANNs have been shown to outperform recent specialised neural network architectures and even strong traditional ML methods. However, ANNs require careful pre- and post-processing to achieve good performance which can be challenging \cite{38}. ANN classifiers are also known for their accurate results on imbalanced datasets where one class is more prevalent in the training data than the another \cite{36}. 

ANNs with considerably more hidden layers (often known as Deep Neural Networks) are typically considered state of the art for many decision based problems in terms of performance, confront various challenges when applied to tabular data compared to white-box models. This includes a lack of localisation and lower performance due to the inner structure that cannot handle all feature types (numerical, ordinal, and categorical)\cite{39}. ANNs often do not perform as well as some white-box models for tabular data\cite{40}. An ANN is considered as a black box model, with little transparency or interpretability into how input data is used in the model’s predictions. Thus, for our study, This model is considered the complexiest and we treat it as a black box model that required extra tools for explanation.

The performance metrics for these models, including sensitivity, specificity, precision, and AUC, the area under the receiver operating characteristic curve, were presented to the workshop attendees as in shown in Table \ref{tbl:performance Metrics}. Sensitivity, which represents a test's ability to correctly identify individuals with a condition, and specificity, which indicates the ability to correctly identify those without the condition, are crucial metrics for evaluating diagnostic test performance. Precision, or positive predictive value, reflects the proportion of true positive results among all positive results, indicating the reliability of a positive test outcome. AUC, is a comprehensive measure of a test's discriminative ability across all possible thresholds, providing a single value to assess overall test performance. These metrics are important for determining the effectiveness  of diagnostic predictions in clinical settings \cite{61}. 

\begin{table}[htbp]
    \centering
    \caption{Performance Metrics for Machine Learning Models Developed to Predict Heart Attack Risk}
    \label{tbl:performance Metrics}
    \begin{tabular}{@{}lllll@{}}
	\toprule
	Model & Sensitivity & Specificity & Precision & AUC\\
        \midrule
	LR & 0.78 & 0.85 & 0.46 & 0.90 \\
	RF & 0.83 & 0.97 & 0.83 &0.96\\
        ANN & 0.75 & 0.85 & 0.43 & 0.80 \\
	\botrule
    \end{tabular}
   \footnotetext{LR: Logistic Regression}
   \footnotetext{RF: Random Forest}
\end{table}

\subsubsection{XAI Methods}
\begin{itemize}
    \item \textbf{Global Explanation}\\

    For each selected ML/AI model we chose a proper method for the global explanation to understand the general structural characteristic. It can be either model-specific method that can only be applied on that specific model or it can be model agnostic method that can be build above any ML/AI model to provide information on its inner working. In global explanations it is common and desired by clinicians to use feature importance/ feature contribution approaches for assigning scores to input features in a predictive model that indicates the relative importance of each feature when making a prediction \cite{49}. The relative scores can highlight which features may be most relevant to the diagnoses. This may be interpreted by a domain expert and can be used as the basis for gathering further data. 
    
   In the logistic regression model, feature importance was assessed using odd ratios, a fundamental measure of logistic regression interpretability. The odd ratio quantifies the change in odds of the outcome for a one-unit increase in a continuous predictor or for one category relative to a reference category in a categorical predictor, assuming other variables remain constant. This attribute makes the model's coefficients interpretable, as they directly indicate the influence of each predictor on the likelihood of the outcome\cite{63}.

The working group was presented with the odd ratios corresponding to the highest-ranked features in the logistic regression model utilised for predicting the risk of heart attack, as shown in Table \ref{tbl:oddratio}. The global interpretation reveals that a history of angina heart attack emerges as a highly significant predictor in the model. Notably, the odd ratio indicates that patients with a history of angina heart attack exhibit around a 53-fold average increase in the odds of experiencing a heart attack compared to those without such a history. Additionally, features such as a history of atrial fibrillation, rheumatoid arthritis, and chronic kidney disease yielded odd ratios of 4, 2.2 and 2 respectively, highlighting their respective contributions to the predictive model. The confidence for these ranking are also appear in the table, showing for angina heart attack for example the true odd ratio lies between 39 and 75, indicating a statistically significant effect since the interval does not include 1. The relatively wide interval suggests some uncertainty around the estimate, but it confirms a strong positive relationship between the predictor and the outcome. and the same interpretation valid for all features.

\begin{table}[ht]
\caption{Feature Importance Ranking Using Odds Ratio in Logistic Regression Model for Heart Attack Risk Predictiont}\label{tbl:oddratio}%
\begin{tabular}{@{}llll@{}}
\toprule
Variables & Odds Ratio  & minimum & maximum\\
\midrule
Angina heart attack    & 53.4   & 39.01  & 75.09 \\
Atrial fibrillation    & 3.95  & 3.19 & 4.92  \\
Rheumatoid Arthritis & 2.24  & 1.44  & 3.52  \\
Kidney Disease & 2.03   & 1.70  & 2.43\\
Region (North East) & 1.98   & 1.51  & 2.68\\
Stroke  & 1.92   & 1.59  & 2.51\\
Hypertension treatment & 1.9   & 0.95  & 3.58\\
Smoking status (Heavy Smoker) & 1.8   & 1.19  & 2.65\\
Diabete (Type 2) & 1.78   & 1.51  & 2.15\\
region (West Midlands) & 1.59   & 1.21 & 1.95\\
\botrule
\end{tabular}
\end{table}

In random forest models, the Gini measure is employed to assess feature importance. Gini impurity quantifies the likelihood of incorrect classification within a dataset, where high impurity denotes a mix of classes and low impurity indicates homogeneity. During the construction of a random forest, data is iteratively split into smaller subsets (via nodes) based on different features, aiming to reduce impurity with each split. The decrease in Gini impurity resulting from each split reflects the influence of the feature used. By averaging the reduction in impurity across all trees in the forest, an overall importance score for each feature is obtained. Features are then ranked according to their average reduction in impurity, with higher scores indicating greater importance \cite{63}. The expert group was presented with Figure \ref{fig:meanDecreaseGini}, illustrating the mean decrease GINI plot for the random forest risk prediction model. Within this model, age emerges as the most influential predictor, boasting a mean decrease in Gini of 780. This indicates that age is  significantly contributing to impurity reduction. Similarly, angina heart attack demonstrates considerable importance in predicting the target variable, with a mean decrease Gini of 730. Weight, systolic blood pressure, and systolic blood pressure STD exhibit mean decreases in Gini around 300. While these values remain relatively high, they suggest that these features exert less impact on impurity reduction compared to age and angina heart attack.

\begin{figure}[htbp]
\centering
    \includegraphics[scale=0.8]{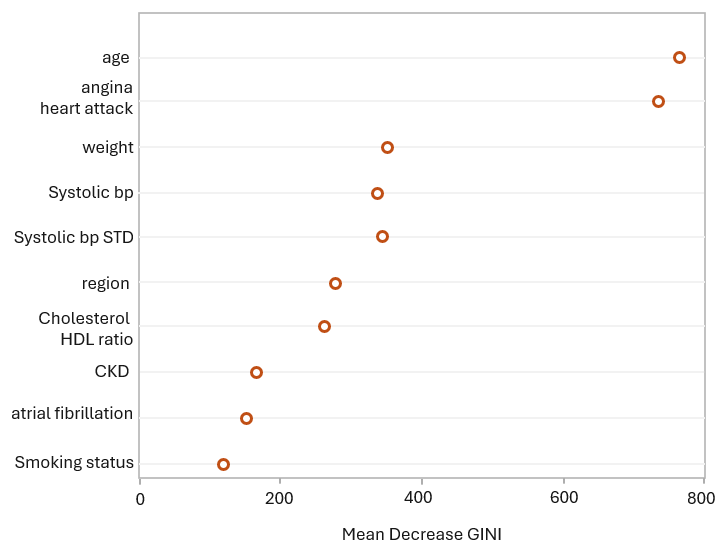}
    \caption{Feature Importance Ranking Using Mean Decrease GINI Metric in Random Forest for Predicting Heart Attack Risk}
    \label{fig:meanDecreaseGini}
\end{figure}

In this study, for the global explanation of the ANN model, a model-agnostic method known as permutation importance was presented in the workshops as a well known method for global explanation in healthcare \cite{47}. Permutation feature importance assesses the impact of each feature on the model's performance by measuring the increase in prediction error when the feature values are permuted while keeping other variables constant \cite{43}. Specifically, a feature is considered significant for the model if permuting its values noticeably increases the prediction error, indicating its importance in the model's predictive performance. Conversely, if the prediction error remains relatively unchanged after permutation, the feature is deemed less useful. Figure \ref{fig:permutationimportance} presents the permutation importance plot for the ANN model, illustrating the decrease in recall (a metric related to sensitivity) following permutation of all features with a confidence bounds. Angina heart attack, with a permutation importance of 0.15, is the most influential feature, suggesting that shuffling its values resulted in the largest increase in prediction error. Age, with a permutation importance of 0.105, follows closely behind in importance. While not as influential as variable angina heart attack, it still significantly impacts the model's predictive performance. Atrial fibrillation, chronic kidney disease and smoking status with permutation importance values of 0.035, 0.025, and 0.015 respectively, have lower importance scores. 

\begin{figure}[htbp]
    \includegraphics[scale=0.8]{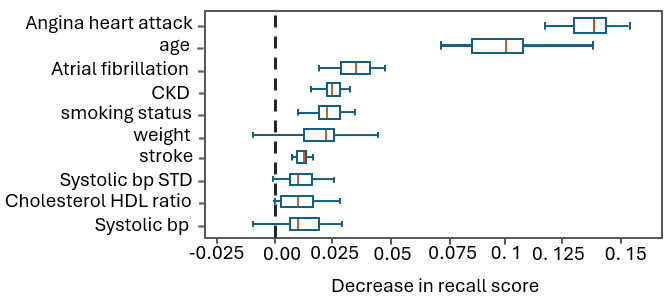}
    \caption{Permutation Importance Ranking of Features for the Neural Network Model in Heart Attack Risk Prediction, Assessing Reduction in Recall Error}
    \label{fig:permutationimportance}
\end{figure}

\item \textbf{Local Explanations} \\
LIME was chosen to obtain local explanations for the ANN model in this study. LIME can be considered as a model-agnostic post-hoc XAI method that provides explanations without opening the complex black box model. To explain a prediction for a specific instance, LIME generates a new dataset consisting of perturbed versions of the instance by slightly altering its feature values. The black-box model is then used to predict outcomes for these perturbed instances, creating a dataset of perturbations and their corresponding predictions. LIME assigns weights to these perturbations based on their proximity to the original instance, with closer perturbations receiving higher weights to emphasise local behavior. An interpretable model, such as linear regression, is then fitted to this weighted dataset, approximating the black-box model's behavior in the local region around the instance of interest. The coefficients of this interpretable model provide the local feature importance, indicating how each feature contributes to the prediction. The resulting explanation highlights the most influential features for the specific prediction \cite{26}. However, LIME is considered unstable due to the randomness in generating perturbed versions of the original instance which can result in different sets of perturbed data for different runs, leading to variability in the explanations. The choice of surrogate model and the specific data points used for fitting can also affect the resulting explanation, making it sensitive to the local sample. Additionally, LIME's weighting scheme, which assigns weights to perturbed instances based on their proximity to the original instance, can introduce instability.

In the workshops, two cases from the test set were used as case studies: a low-risk heart attack case misclassified by the model and a high-risk heart attack case correctly classified. To address stability and ensure consistent explanations, LIME was run 20 times for each example. Figures \ref{fig:FB_ime} and \ref{fig:TP_lime} show LIME local explanations for the two cases, respectively. Each figure provides the prediction confidence (prediction probability). In addition, it lists features on one axis, with bars indicating their importance; the direction of each bar shows whether the feature contributes positively or negatively to the prediction, while the length indicates the strength of this contribution. Feature values for each instance contextualise their importance. LIME provides local explanations specific to the examined instance, and feature importances can vary across predictions. Comparing the explanations for the two cases reveals the model's consistency in using the same features, with angina heart attack as the most important feature, followed by age, atrial fibrillation, and rheumatoid arthritis. In the misclassified case, the patient was older than 75 and had a history of atrial fibrillation, leading to a high-risk classification despite not being high-risk. In the correctly classified case, having an angina heart attack and being older than 75 led to correctly identify the patient as high-risk.

\begin{figure*}[htbp]
   \captionsetup{font=footnotesize,labelfont={bf,sf}}
   \hspace{-2em}
     \begin{subfigure}[b]{0.5\textwidth}
         \centering
         \includegraphics[scale=1]{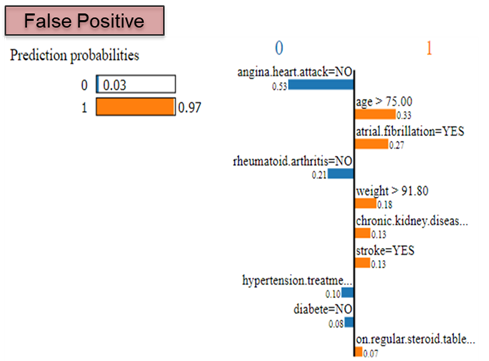}
         \caption{False positive case}
         \label{fig:FB_ime}
     \end{subfigure}
     \hfill
      \begin{subfigure}[b]{0.5\textwidth}
         \centering
         \includegraphics[scale=1]{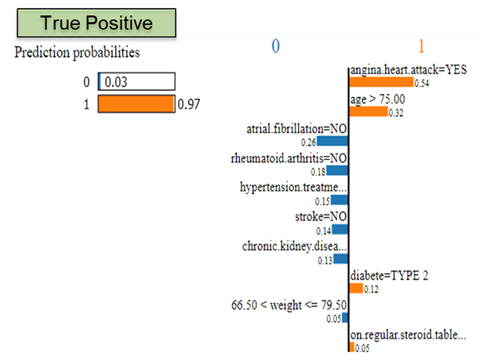}
         \caption{True positive case}
         \label{fig:TP_lime}
     \end{subfigure}
        \caption{LIME local explanations for two selected cases predicted by the neural network model: (a) local explanation for a false positive output, (b) local explanation for a true positive output.}
        \label{fig:lime}
\end{figure*}

\item \textbf{Counterfactuals} \\

For counterfactual generation, we used the Explainable Matrix "ExMatrix" technique to extract counterfactual explanations from random forest. ExMatrix identifies decision paths within the random forest that lead to different predictions and searches for the nearest path resulting in an alternate decision, such as moving from "low risk" to "high risk." The method focuses on identifying the minimal feature changes required to switch decision paths, providing clear guidance on which features need to be modified and by how much \cite{64}.

Both the original version of ExMatrix, as introduced by \citep{64}, and a simplified version was presented to the expert group to assess their ability to interpret the complexity of the visualizations. This dual presentation allowed us to evaluate how the complexity of data presentation impacted interpretation. Simplification was achieved by aggregating the feature-wise changes required across all decision trees in the random forest, deriving a single value for each feature that represents the extent to which modifying that feature would alter the prediction.

Figure \ref{fig:exmatrix} shows the original and the simplified version of the ExMatrix counterfactual visualisation for a selected true positive case. In the modified version, the yellow bars indicate negative alterations and the green bars positive ones needed to change a high-risk case to low-risk. For example, transitioning a high-risk patient with angina, type 2 diabetes, and erectile dysfunction to low-risk involves removing these conditions, particularly focusing on treating angina, type 2 diabetes and  erectile dysfunction and theoretically reducing the patient's age.

\begin{figure}[htbp]
   \captionsetup{font=footnotesize,labelfont={bf,sf}}
   \hspace{-4.5em}
     \begin{subfigure}[b]{0.69\textwidth}
         \centering
         \includegraphics[scale=1]{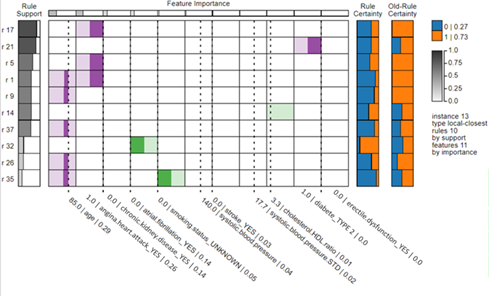}
         \caption{Original ExMatrix counterfactual}
         \label{fig:OrigExmatrix}
     \end{subfigure}
     \hfill
      \begin{subfigure}[b]{0.35\textwidth}
         \centering
         \includegraphics[scale=1]{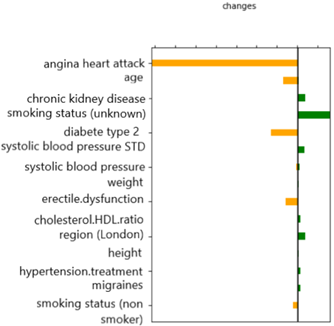}
         \caption{Simplified ExMatrix}
         \label{fig:simplifiedExmatrix}
     \end{subfigure}
        \caption{Exmatrix counterfactual cases for a true positive case. Figure (a), Original matrix visualisation . Figure (b), simplified 1D plot of ExMatrix.}
        \label{fig:exmatrix}
\end{figure}

\end{itemize}

\subsection{Results \& Expert Working Group Insights}
In this study, we presented the AI/ML models employed to predict heart attack risk to the expert group. Performance metrics for these algorithms were subsequently shared with them, as detailed in Table \ref{tbl:performance Metrics}. All models demonstrated acceptable performance, but random forest outperformed the others. Sensitivity, essential for accurately identifying high-risk cases, was particularly stronger for random forest. However, regulators expressed concerns about the selection of predictive models for CDSS. While complex models like neural networks may initially seem attractive, regulators emphasise the importance of interpretability and ease of understanding to ensure patient safety. Simpler models, such as logistic regression or simple random forest, might be more suitable if they offer comparable performance, as they are easier to understand and interpret. Comprehensive research is needed to evaluate the trade-offs between model complexity and interpretability in the intended clinical settings.

Clinicians on the other hand expressed their interest in the overall model performance and ensuring regulatory approval rather than delving into the technical aspect of the AI/ML models. 

All the global explanation methods presented in the workshops utilised feature ranking to determine the most influential features on ML/AI models' predictions. Table \ref{tbl:feat rank} compared the top five features across the three chosen models. Commonly ranked features included angina, age, and atrial fibrillation, but there were clear differences between models. For regulators, this variation in influential features across different models highlighted the necessity of assessing multiple ML/AI models before selecting one for AI-CDSS. The assessment needs to consider not only performance metrics but also the features affecting model predictions. Regulators also highlighted the importance of evaluating how these features related to the specific use case and aligned with clinical knowledge. Which require comprehensive testing and active involvement of clinicians to identify the most suitable model. This also aligned with clinicians' preference for being informed about the underlying considerations and features that AI-CDSS relied on for their predictions, expressing that this transparency is essential for them to build trust in the CDSS system's outputs.

\begin{table*}[htbp]
		\centering
  \caption{The top five important features ranked from the global explanation methods}
		\label{tbl:feat rank}
		\begin{tabular}{@{}llll@{}}
			\toprule
            Rank & Logistic Regression & Random Forest & Neural Network\\
		\midrule
			1 & angina heart attack & age & angina heart attack \\
			2 & atrial fibrillation  & angina heart attack & age\\
             3 & rheumatoid arthritis  &  weight & atrial fibrillation \\
             4 & chronic kidney disease & Systolic blood pressure & chronic kidney disease \\
             5 & region (North East) & Systolic blood pressure STD & smoking status \\
			\botrule
		\end{tabular}
	\end{table*}

In the LIME local explanations presented in Figure \ref{fig:lime}, attributes were ranked by their influence on heart attack risk, along with the model's confidence score (prediction probability) for each class. The ANN model misclassified the first example (Figure \ref{fig:FB_ime} a false positive case) but correctly predicted the second (Figure \ref{fig:TP_lime} a true positive case). In both cases, the model was 97\% confident in its predictions.

For the correctly classified case, this high confidence was seen by both clinicians and regulators as a way to enhance clinicians' confidence in their predictions. However, the discrepancy between model confidence and prediction error in the false positive case was highlighted as a significant issue. The explanation revealed that the false positive classification was due to the patient being older than 75, having a history of atrial fibrillation, kidney disease, and stroke, and a weight above 91.80 kg which are clinically valid explanations but still flagged an issue. While this specific false positive case may not pose an immediate issue, a high-confidence misclassification of a high-risk case as low risk would raise significant safety concerns for both clinicians and patients. 

 Another point raised by the clinicians was that the LIME visualizations were not easy for them to interpret, even though the data science team found them simple. Clinicians argued that while LIME outputs show how each feature contributes to the final prediction, in clinical settings, these visualizations would require significant time to interpret and evaluate how combinations of features diagnose a patient as high or low risk. This complexity could hinder their practical use in fast-paced clinical environments, where quick and accurate decision-making is crucial.

The original visualization for ExMAtrix counterfactual explanation, represented in Figure \ref{fig:OrigExmatrix}, was initially considered informative as it presents the decision paths for all trees in the random forest. However, the expert group found it to be complex and overwhelming for use in clinical settings, especially with a significant number of trees in the model. In its simplified form (Figure \ref{fig:simplifiedExmatrix}), the counterfactual cases were deemed easier for clinicians to interpret. Despite this improvement, the clinicians noted that, similar to LIME, it would still require considerable time to utilize the output effectively in a clinical environment. There was a consensus among the experts that certain counterfactual features, such as age, gender, and some features of medical history, are difficult or impossible to change. Interpreting these features in a clinical setting would be overwhelming and would not necessarily guarantee a change in the model's output if those features were excluded from the required changes to alter the risk prediction.

\section{Stage 2: Pilot Study}
In this stage, we conducted a pilot study to evaluate the impact of AI/XAI methods and visualizations on clinicians' diagnostic processes. Clinicians, who were not part of the expert group, were presented with the same XAI outputs as outlined in section \ref{sec:methods}. This investigation sought to evaluate the influence of these outputs on their diagnostic processes and to investigate the key considerations and challenges entailed in clinicians' engagement with explanations in CDSS. The results of this pilot study were then discussed by the expert group. The aim was to investigate both the actual results of the pilot study and the expert group's reflections on these findings.

In accordance with the Health Research Authority (HRA) guidelines for clinical research within the UK's National Health Service (NHS), ethical approval was not required for this study. We used the HRA's decision tool to confirm this, and the tool's decision is provided in Appendix A.

\subsection{Methods}

This study involved several stages. Initially, the data science team selected fifty synthetic patient records from the same test dataset used in earlier phases of the research, ensuring they were representative of the AI/ML model’s performance and covered a range of patient demographics. The clinical team then refined this selection to ten patient records, focusing on cases that would provide diverse and clinically interesting scenarios.  The cases chosen consisted of four true positives (correctly classified as high-risk), four true negatives (correctly classified as low-risk), one false negative (misclassified as low-risk), and one false positive (misclassified as high-risk). This selection allowed the expert group to examine a diverse range of decision scenarios. 

These records were presented to eight practicing clinicians, ensuring that they were provided with exactly the same data that has been used by the AI/ML models and consists of patients medical history. These clinicians were shown each patient’s full medical history individually, one patient at a time. For each patient case, clinicians were first asked to make their own diagnoses based solely on the patient's medical data and identify the top features that influenced their decision-making process.

Once they had completed their diagnosis for a given patient, they were shown the corresponding AI diagnosis, along with explanations from the XAI models and the confidence level of the CDSS predictions. This allowed the clinicians to assess whether and how the AI/XAI outputs might influence their diagnostic decisions and confidence levels. The process was repeated for all ten patients in sequence — clinicians reviewed the data, made their diagnosis, reviewed AI/XAI outputs, and then moved to the next patient.
 
Finally, the outcomes of this study, including both the clinicians' diagnoses and their responses to the AI/XAI outputs, were presented to an expert group for further discussion and analysis.

\large\textbf{3.2 {Pilot Study Results \& Regulatory Reflection on the Results}}

To evaluate the pilot study results, we began by comparing the clinicians' diagnoses with the output from the ML models. Specifically, we assessed clinician confidence in their diagnoses prior to seeing the ML outputs and compared it to the confidence of the ANN model.  Clinical confidence was measured by the percentage of clinicians diagnosing each case as high-risk. Confidence in high-risk diagnoses is complementary to that in low-risk diagnoses, meaning a lower confidence in high-risk implies a higher confidence in low-risk and vice versa. Confidence levels below 50\% are considered low-risk diagnoses, while levels of 50\% or higher are considered high-risk diagnoses.

\begin{figure*}[htbp]
\centering
    \includegraphics[scale=0.65]{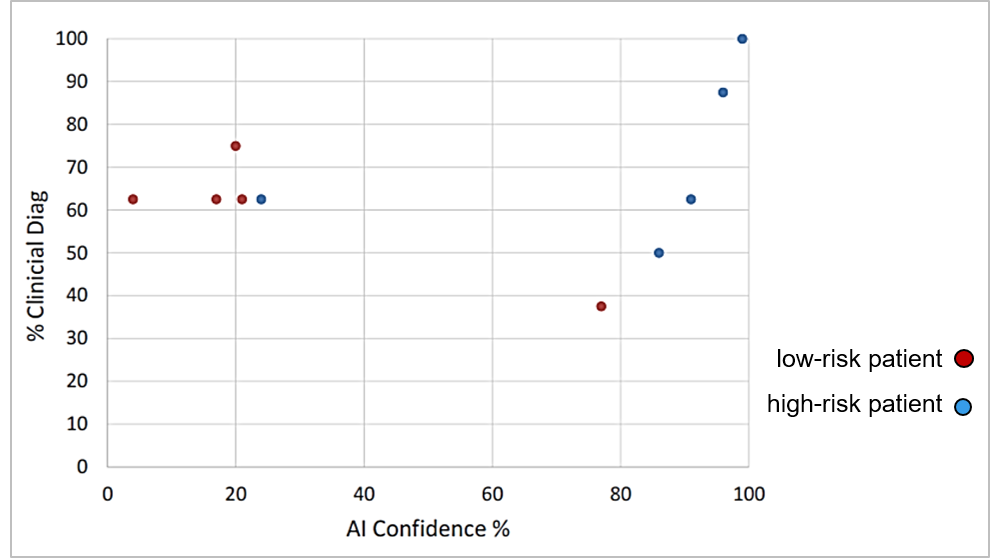}
    \caption{Clinical Confidence Vs ANN model Confidence in Diagnosis Patients as High Risk to Have a Heart Attack}
    \label{fig:AIvsclinical-conf}
\end{figure*}

Figure \ref{fig:AIvsclinical-conf} represents this comparison, with the ten cases color-coded according to their actual risk status. Ideally, high-risk patients would cluster at the top right of the plot, and low-risk patients at the bottom left, reflecting high confidence in accurate diagnoses. The ANN model's confidence was either very high or very low, correlating with the actual risk status of the patients, while clinician confidence tended to cluster in the middle, showing a tendency for false positives as clinicians often rated more cases as high-risk.

Clinicians and the AI model agreed on 4 out of 5 high-risk cases, accurately diagnosing them as high-risk. Therefore, clinicians succeed in diagnosing all five high-risk cases being as high-risk. For the low-risk cases, the AI model correctly identified four, but only one was correctly diagnosed as low-risk by the majority of clinicians. This particular low-risk case was also the only negative case that the AI misdiagnosed.

For regulators, the disagreement between the actual risk status, the output of AI/ML models, and clinicians' diagnoses highlighted a potential area for consideration, suggesting that the disagreement may not necessarily be problematic. Specifically, for actual low-risk patients who were diagnosed as high-risk by clinicians and/or models (false positive cases), this discrepancy could be attributed to incomplete data coverage of the patients' full medical histories. It is possible that these patients might have experienced a heart attack after the data collection period, indicating that the model's and clinicians' high-risk diagnoses could be justified despite the initial low-risk status.

In assessing influential features, there was generally strong agreement between clinicians and AI/ML models on high-ranking features such as age, previous angina, type 2 diabetes, smoking, and cholesterol. However, a notable divergence emerged for other features, with AI models incorporating a wider range of variables that did not consistently align with clinicians' considerations. Regulators suggested that this divergence could make AI models a complementary tool, providing a new perspective during the decision-making process. 

In assessing influential features, there was generally strong alignment between clinicians and the AI/ML models on key features such as age, history of angina, and type 2 diabetes. This assessment compared the top 3 features identified by clinicians with those highlighted by LIME explanations after running LIME 10 times. Features appearing at least once in the top 3 in LIME were included for comparison.

\textbf{For the high-risk cases that were accurately diagnosed by both the ML model and the clinicians:}

\textbf{Patient 1:} The ML model ranked diabetes, angina, and weight as the top features. Among clinicians, 5 out of 8 included diabetes in their top 3, 6 ranked angina, and 2 ranked weight.

\textbf{Patient 2:} The ML model identified angina, atrial fibrillation (AF), and weight as the most important factors. Here, 7 of 8 clinicians included angina in their top 3, 2 included AF, and 4 ranked weight.

\textbf{Patient 3:} The ML model ranked diabetes, angina, and AF as the key features. In this case, 4 of 8 clinicians included diabetes and angina in their top 3, 3 included AF, and 4 ranked weight.

\textbf{Patient 4:} The ML model highlighted angina, age, and weight, and 4 clinicians included these features as the most influential features in their diagnosis.

Other variables mentioned by fewer clinicians in these high-risk cases included smoking status and Chronic Kidney Disease (CKD), which the ML model typically ranked lower, in the 4th and 5th positions. Additionally, the ML model sometimes included features like stroke or systolic blood pressure that clinicians did not prioritise.

\textbf{For the patients whom clinicians diagnosed as high risk, but the ML model correctly classified as low risk:}

\textbf{Patient 6:} Five clinicians ranked weight as a top 3 feature, consistent with the ML model. 5 clinicians cited diabetes, and 4 considered the patient's status as an ex-smoker to be key. However, the ML model aligned with only 2 clinicians, focusing on age and systolic blood pressure as the main factors for classifying this patient as low risk.

\textbf{Patients 7 and 9:} In both cases, 3 and 6 clinicians, respectively, agreed with the ML model that diabetes was a top feature. However, 6 clinicians considered slightly elevated cholesterol and erectile dysfunction to be significant risk factors, leading them to classify the patients as high risk. whereas the ML model did not prioritize these factors, instead focusing on the absence of a history of angina as key to assessing low risk.

\textbf{Patient 8:} The ML model correctly predicted the patient as low risk due to the lack of a history of heart disease (AF, angina) or diabetes, despite the patient's advanced age (84). However, 5 of 8 clinicians saw the history of stroke as a significant risk factor, and 4 considered slightly elevated hypertension important in their diagnosis.

\textbf{Cases Misclassified by the ML Model:}

For the patient incorrectly predicted as high risk by the ML model but correctly identified by clinicians, the model ranked age (over 75) and CKD as important features. Three clinicians who also misdiagnosed the patient as high risk cited these same factors. The remaining clinicians correctly diagnosed the patient, citing a clear medical history and stable metrics (hypertension, blood pressure, and cholesterol) as indicators of low risk.

In the case of a patient falsely predicted as low risk by the ML model but correctly diagnosed by most clinicians, 4 clinicians cited erectile dysfunction, diabetes, and slight overweight as risk factors. While the ML model agreed on the importance of diabetes, it downplayed these factors, instead focusing on the absence of angina, AF, or CKD, leading to the low-risk classification.

From a regulatory perspective, the observed alignment between clinicians and the ML model explanations regarding major risk factors, such as diabetes, angina, and weight, indicates that the model effectively identifies critical clinical indicators associated with high-risk cases. This consistency suggests that the model has the potential to meet safety requirements for identifying high-risk patients, provided that these factors are rigorously validated against clinical standards and that the model's performance aligns with its intended purpose. Furthermore, the model’s emphasis on the absence of certain conditions (e.g., angina or atrial fibrillation) as protective factors has resulted in several accurate low-risk predictions where human clinicians may have misjudged the risk. While this could imply enhanced performance in specific scenarios, it underscores the necessity for careful evaluation of how the model balances the absence of conditions with existing risk factors. Clinicians also often considered traditional markers, such as cholesterol levels and erectile dysfunction, when assessing risk, even though these factors are not consistently prioritized by the ML model. This disparity indicates that clinicians may give greater weight to patient history, while the model is primarily driven by data patterns. In borderline cases (e.g., Patients 6, 7, and 9), notable discrepancies between the model's reasoning and the clinicians’ decisions became apparent, highlighting potential gaps in the model’s training. Nevertheless, these differences may also underscore the value of models in providing clinicians with new perspectives, albeit requiring rigorous testing to ensure their reliability and effectiveness.

After revealing AI diagnoses, explanations, and decision confidence, all clinicians agreed that these explanations boosted their confidence in their diagnoses, particularly when the explanations and AI predictions aligned with their clinical knowledge. In cases where discrepancies existed between AI and clinician diagnoses, the introduction of XAI  outputs; including influential features and AI confidence, prompted a notable shift in clinicians' diagnostic decisions. Specifically, in five out of six instances, clinicians adjusted their diagnoses to align with AI prediction, resulting in an overall improvement in diagnostic accuracy, especially concerning low-risk cases. However, a notable exception occurred when the AI inaccurately diagnosed a high-risk case, prompting clinicians to adopt a wrong diagnosis in alignment with the AI output. Regulators viewed this as an indicator of automation bias or trust calibration issues.

\section{Discussion}
The goal of this study  was to evaluate and understand multidisciplinary perspectives on the use of AI/XAI technologies in clinical settings and ensure their safe introduction. Initially, we presented ML/AI models with different levels of complexity to regulators and clinicians to observe their reactions. We also introduced global and local explanations for these models to assess how utilising these methods could help clinicians and identify potential challenges for both regulatory and clinical applications. While there are a few studies that examine clinicians' perspectives on using AI \cite{49,54,65} and others that consider regulating AI systems in healthcare\cite{57,59}, our approach is unique in facilitating comprehensive discussions among various stakeholders through workshops. This approach allowed for a better understanding of the needs and concerns of each group.

In the second part of the work, we conducted a pilot study to evaluate human-clinician interaction with AI/XAI systems. Similar to other studies, we assessed the performance of clinicians before and after introducing the XAI outputs. Consistent with previous findings, we observed an overall increase in diagnostic accuracy after incorporating XAI explanations \cite{12,47}. However, incorporating AI in the decision-making process raised potential issues previously noted in similar experiments, such as trust calibration problems and over-reliance on AI system outputs \cite{12,49, 65, 66}. Unlike prior studies, our research included regulators to assess these issues from a regulatory standpoint.

In this section, we discuss the lessons learned from this study and propose regulatory recommendations to consider for safely utilising AI/XAI in clinical settings. These recommendations aim to address the identified challenges and ensure that AI technologies improve clinical decision-making without compromising safety.
\\

\textbf{Lesson 1:  Assessing scientific and analytical validity is essential for AI-CDSS adopting} \\
Before adopting AI/ML based CDSS, it is essential from a regulatory perspective to ensure thorough validation. This is particularly important when using black-box AI systems that lack straightforward scientific and analytical validation methods, especially in high-risk clinical scenarios. Key performance metrics such as sensitivity, precision, and specificity must be rigorously assessed. Additionally, it is important to guide analytics teams to build models that are as simple as possible while still achieving the required tasks \cite{14,23}.
Engaging clinicians for expert review, conducting clinical trials, and implementing XAI  methods are crucial to ensure clinical relevance and transparency. Ensuring regulatory compliance, continuously updating model training, and establishing ongoing monitoring and feedback mechanisms are essential steps to maintain safety in clinical settings.\\

\textbf{Lesson 2: The divergent importance of global explanations for regulators and clinicians}

When approving or adopting AI models, global explanations hold differing significance for regulators and clinicians. For regulators, global explanations of AI/ML models were crucial as they provided insights into how the model functions and what considerations are being made. This understanding is essential for ensuring the model's safety and transparency. Regulators also found value in using feature importance rankings to identify key features that influence the models' predictions. Notably, the variation in feature rankings among different models can be insightful, suggesting that models may learn differently and could be relevant for different scenarios and tasks. This variation should be considered in future research, despite some authors \cite{68,69} suggesting that inconsistent explanations across models might be an issue and invalidate their use.

Clinicians on the other hand, were primarily interested in understanding the clinical knowledge that has been incorporated into predicting patient outcomes. However, they were less concerned with the in-depth technical details of the model's inner workings. Clinicians expressed that they prefer a straightforward explanation of these considerations, ideally provided in a concise briefing before they begin using the AI/ML-based CDSS. \\

\textbf{Lesson 3: Local explanations are essential for clinicians but demand careful consideration in both method selection and usage}: \\
 Clinicians in the workshop and pilot study found local explanations beneficial for enhancing their trust in AI model decisions, particularly when the explanations and key features aligned with their clinical knowledge. 
 
 Regulators, did not show a strong interest in understanding how a decision for a specific case is made. Their primary concern was ensuring that human clinicians can interact with these explanations in a safe and effective manner. Consequently, a brief overview on how to interpret XAI methods' outputs was provided to clinicians before the experiment in the pilot study. This preparation was reflected in their satisfaction and their ability to understand the outputs. In our study we used LIME as local explainer because its output is similar to how humans clinicians visualise explanations \cite{53}. However, in a clinical environment, this might not be the best method. Clinicians in the workshops expressed that the visualizations can be confusing and time-consuming in a fast-paced clinical setting, especially when practitioners need to assess from the visualisation how a combination of features contributed to a specific diagnosis. \\

\textbf{Lesson 4: Model confidence is a key for trust and safety but can cause issues in some scenarios}\\
While confidence is considered as a reliable indicator of the degree that clinicians can trust the output of a machine learning model, it is vital to devote careful attention to the basis of this measure as well as any inherent limitations (missingness and bias) in the training data. This includes factors like the severity of specific medical conditions, extra symptoms and conditions that the treated patient may be suffering and that were not included in the training data. As a result, practitioners must be aware of the data used for the development of AI systems to be able to better understand the decisions being recommended to them and accept or reject these based on this understanding. Special focus should be placed on the correlation between AI model confidence in its predictions and the accuracy of these predictions as was seen in the false positive case in the lime explanation output. There might be trust implications if the model confidence does not match the likelihood of the decision being correct. In such circumstances, an investigation into the potential causes of being \textit{incorrect but confident} should be conducted. \\

\textbf{Lesson 5: Transparency requirements are different for different stakeholders}:\\ 
Regulators in the workshops highlighted the importance of focusing not just on assessing the overall performance of the model, but also, its fairness and any potential biases. This evaluation involves identifying situations where the model demonstrates sub-optimal performance and determining which subgroups are disproportionately affected by these shortcomings. When applicable, documentation should explicitly highlight the affected subgroups and outline situations where the model fails to operate as intended.

Meanwhile, clinicians prioritised understanding the data and clinical considerations during the AI/ML based CDSS development over the actual working logic of the models deployed. This is to ensure that the model assumptions and parameters align with clinical knowledge. However, all workshop attendees including clinicians assured the importance of educating healthcare providers on when to utilise these models effectively and when it may be prudent to avoid their use due to concerns about fairness and safety.\\

\textbf{Lesson 7: Counterfactual explanations are highly useful if they were introduced correctly} \\ Counterfactual analysis is regarded as one of the most important tools in clinical healthcare by experts \cite{51}. However, in a clinical setting, it is critical to communicate these counterfactual scenarios  in a manner that is easily comprehensible for end users, highlighting the main features to change and their related values clearly. Furthermore, considering the nature of the clinical field, experts should be provided with a variety of feasible options. Options, for example, should not recommend modifications to fixed patient characteristics such as age, gender, or medical history as it was introduced in the ExMatrix output.

\subsection{Recommendation for safely introducing AI/XAI tools to clinical settings}

Before adopting AI/ML based CDSS, multiple factors must be considered to ensure their safety and efficacy in clinical workflows. It is generally preferable to use simple yet efficient models as has been suggested by \cite{43}, balancing complexity with interpretability, as simple models are often easier to understand and trust, which is crucial in clinical settings. During model evaluation, it is important to recognize the limitations of the test data, particularly the time span and scope of the patients' medical histories it covers. Assessing multiple models is essential in the selection process, and global explanations should be examined to identify the most influential features for each model. Clinical knowledge should be engaged to evaluate the relevance of the top-ranking features for each model to the specific use case, ensuring the chosen model aligns well with clinical needs and expectations.

Clinicians' preferences for explanation methods and types vary significantly and often differ from developers' preferences \cite{67}. Therefore, involving a diverse group of clinicians from different backgrounds and experience levels in the development process is crucial. For example, while developers might find LIME to be straightforward, some clinicians may find these explanations confusing. Incorporating insights from psychologists and cognitive specialists can also enhance the design process by ensuring that explanations are tailored to the cognitive needs of end-users \cite{12}. Local and counterfactual XAI methods, are encouraged to be used as valuable educational tools, particularly for junior doctors, helping bridge the gap between theoretical knowledge and practical application, especially if their performance has been validated against actual clinical outcomes. Involving experienced clinicians in validating these tools can further enhance their educational value.

AI and XAI tools should be regarded as support systems rather than standalone solutions \cite{59}. They are most effective when complementing human practitioners, highlighting information that might otherwise go unnoticed. To facilitate this complementary relationship, professional training and education for all stakeholders are essential. This training ensures that medical professionals are prepared for AI integration, understand how to calibrate their trust in AI outputs, and know when to rely on or discard CDSS recommendations.

In our workshop, the inclusion of perspectives from psychologists and cognitive specialists could provide deeper insights into how clinicians prefer to be provided with AI/XAI outputs. Additionally, employing other well-known methods for local explanations, such as SHAP \cite{27}, and counterfactual methods like DiCE, which allow restrictions on counterfactual cases, could be beneficial. This is particularly relevant in clinical settings where factors such as demographics and medical history cannot be changed. Future work will include a pilot study with a broader scope to assess how clinicians' experience levels and background knowledge of AI affect their interaction with AI-CDSS. Observing practitioners' interactions with these systems over a longer period will also be beneficial to address potential challenges and opportunities in AI/XAI medical devices.


\section{Conclusion}

This  study evaluated  the perspectives of data scientists, clinicians, and regulators regarding the safe integration of ML/AI-based CDSS into clinical settings. Introducing XAI methods as a crucial tool for ensuring the safe deployment of these technologies. We found that performance metrics, in conjunction with global explanations and clinical knowledge, serve as valuable guides for selecting suitable AI models for specific tasks. Local explanations are essential for improving clinician trust. However, Regulators underscored the significance of viewing AI/XAI CDSS as support systems only and emphasized the need for proper trust calibration to mitigate potential risks such as over-reliance on AI outputs.

 \backmatter





\bmhead{Abbreviations}
AI: Artifcial Intelligence; ML: Machine Learning; ANN: Artificial Neural Network; CDSS: Clinical Decision Support System; XAI: eXplainable AI; MHRA: Medicines and Healthcare products Regulatory Agency; AI-CDSS: AI based Clinical Decision Support System; LIME: Local Interpretable Model-agnostic Explanations; ExMatrix: Explainable Matrix; SHAP: Shapley Additive explanations; DiCE: Diverse Counterfactual Explanations

 \section*{Declarations}

\begin{itemize}
 \item Ethics approval and consent to participate: Not applicable
 \item Consent for publication: Not applicable
 
 \item Availability of data and materials: CPRD cardiovascular disease synthetic dataset used in this paper can be requested from CPRD (https://cprd.com/cprd-cardiovascular-disease-synthetic-dataset)
 
 \item Competing interests: Not applicable 
 
 \item   Funding: This work was funded by the Regulators Pioneer Fund 3, Department for Science, Innovation and Technology. The RPF is a grant-based fund to enable UK regulators and local authorities to help create a UK regulatory environment that encourages business innovation and growth. The current £12m round is being delivered by DSIT. This work was also supported by the UK Regulatory Science and Innovation Networks – Implementation Phase: Human Health CERSIs programme through the project RADIANT: Regulatory Science Empowering Innovation in Transformative Digital Health and AI (Grant Ref: MCPC24031), funded by the Medical Research Council (MRC) and Innovate UK.. 
 
\item  Authors' contributions: P.M. and R.B. initiated the project and conceived the study. A.T. guided the overall project and experiments, providing essential insights for interpreting the results and discussing their implications. H.A. offered clinical guidance for the pilot study and selected the patients. D.A., A.K. and A.T. performed the analysis, conducted the experiments, and designed the pilot study. D.A. was the primary manuscript writer. All authors contributed to revising the initial draft and approved the final version of the manuscript.

 \item Acknowledgements:  Many thanks to the expert group that consisted of: Akish Luintel (UCLH), Alan Davies (Health Education England), Allan Tucker (Brunel), Andras Meczner (Live Healthily), Asal Khoshravan Azar (Brunel), Dima Alattal (Brunel), Emma Keeling (Imperial College), Hatim Abdulhussein (NHS England), Johan Ordish (Roche), Maria Wilder (BEIS), Melissa Tucker (Health Education England), Michael Nix (Leeds NHS Trust), Paul Campbell (MHRA), Puja Myles (MHRA), Richard Branson (MHRA), Russell Pearson (MHRA), Salah Hammouche (Health Education England), Susan Hodgson (MHRA).

 \end{itemize}


\bibliography{sn-article}
\newpage
\section*{Appendix A: Ethical Approval Requirements for the Pilot Study)}

\begingroup
\sbox0{\includegraphics[scale=0.9,page=1]{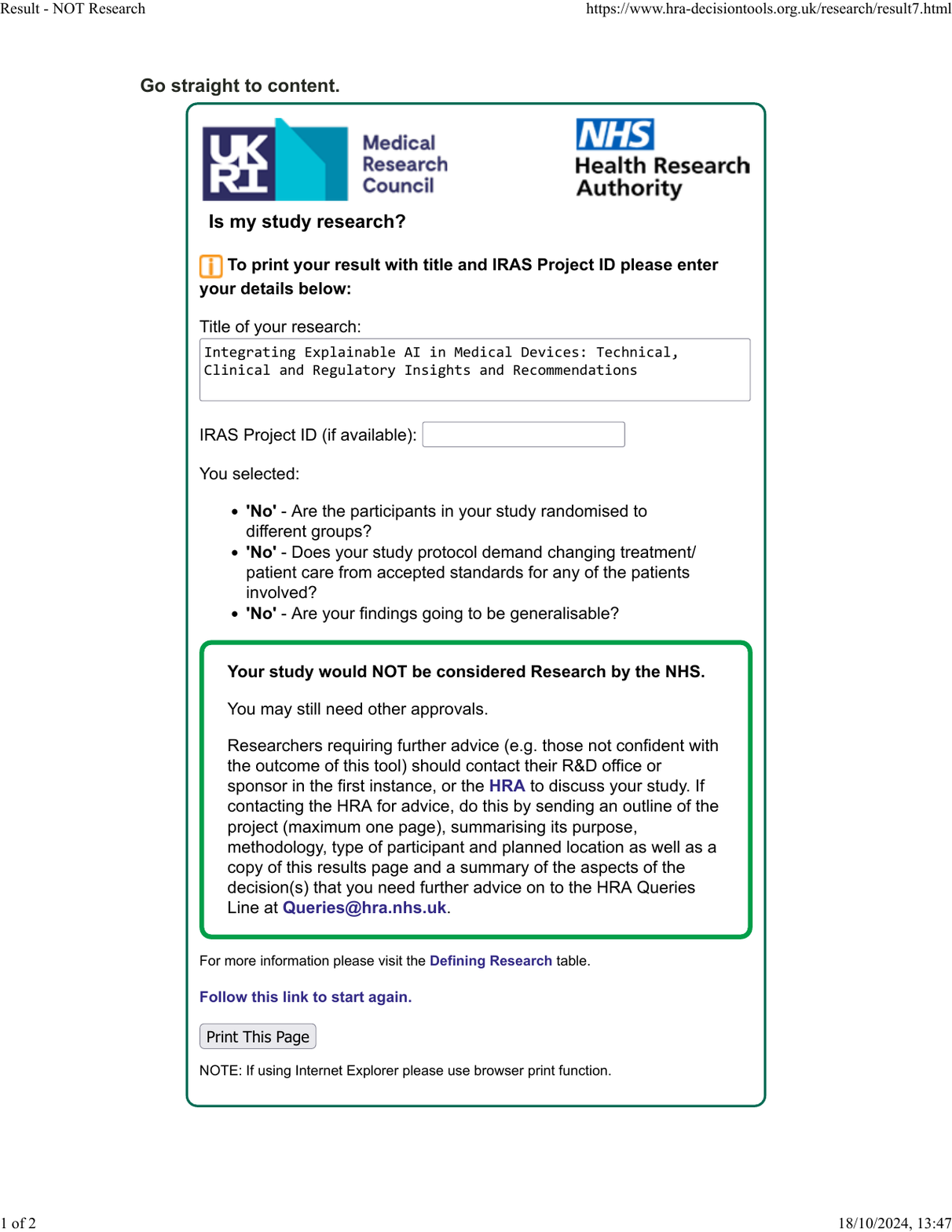}}%
\includegraphics[scale=0.9,page=1, clip,trim={.2\wd0} 0 {.2\wd0} {.12\wd0}]{ExplainabilityEthics.pdf}
\endgroup

\end{document}